# Electric-Field-Induced Resistive Switching in a Family of Mott Insulators : towards Non-Volatile Mott-RRAM Memories

By *Laurent Cario[#], Cristian Vaju, Benoit Corraze[#], Vincent Guiot, Etienne Janod[#,*]*

Institut des Matériaux Jean Rouxel (IMN), Université de Nantes, CNRS, 2 rue de la houssinière, BP 32229, 44322 Nantes Cedex 3, France

[*]   Corresponding-Author
E-mail: Etienne.janod@cnrs-imn.fr

[#] these authors contributes equally to this work

The fundamental building blocks of modern silicon-based microelectronics, such as double gate transistors in non-volatile Flash memories, are based on the control of electrical resistance by electrostatic charging. Flash memories could soon reach their miniaturization limits mostly because reliably keeping enough electrons in an always smaller cell size will become increasingly difficult [1]. The control of electrical resistance at the nanometer scale therefore requires new concepts, and the ultimate resistance-change device is believed to exploit a purely electronic phase change such as the Mott insulator to insulator transition [2]. Here we show that application of short electric pulses allows to switch back and forth between an initial high-resistance insulating state ("0" state) and a low-resistance "metallic" state ("1" state) in the whole class of Mott Insulator compounds $AM_4X_8$ (A = Ga, Ge; M= V, Nb, Ta; X = S, Se). We found that electric fields as low as 2 kV/cm induce an electronic phase change in these compounds from a Mott insulating state to a metallic-like state. Our results suggest that this transition belongs to a new class of resistive switching and might be explained by recent theoretical works predicting that an insulator to metal transition can be achieved by a simple electric field in a Mott Insulator [3,4,5]. This new type of resistive switching has potential to build up a new class of Resistive Random Access Memory (RRAM) with fast writing/erasing times (50 ns to 10 µs) and resistance ratios $\Delta R/R$ of the order of 25% at room temperature.

The huge non-volatile memory market is led by the Flash technology, used *e.g.* in Flash cards and Solid State Drives. However, the intrinsic drawbacks of this technology (long writing/erasing time : 1 µs / 1 ms ; use of high voltages : 12 V; complex design)[1] will limit its development in the near future. Resistive Random Access Memories (RRAM) are currently considered to overcome these limitations. In RRAM, information storage is enabled by a non-volatile and reversible switching of a resistive device between two different resistance states. This resistive switching is obtained by the simple application of short electric pulses to an active material. A large variety of materials are known to exhibit a resistive switching (RS) phenomenon, such as transition metal oxides (NiO, $TiO_2$, $SrTiO_3$, manganites …) or copper/silver based chalcogenides. So far, different mechanisms based on thermochemical or electrochemical effects have been proposed to explain the non-volatile RS observed in these materials [6]. For example, in the Cu/Ag based chalcogenides, like $Ag_2S$,[6] or in oxides like $SrTiO_3$ [7], the RS is related to cation or anion electromigration, leading either to metallic Cu/Ag dendrites or metallic paths related to the Ti valence change. On the other hand, the RS in $Ni_{1+x}O$ is associated to a thermochemical effect, *i.e.* the formation or destruction of Ni conductive filaments by Joule heating [8]. Interestingly all these mechanisms proposed for RRAM involve local Chemical modifications. Conversely, RRAM based on an Electronic Phase Change are still experimentally unexplored, while they could offer more reliability and higher speed [6,7]. In that respect, the recent theoretical prediction that a simple electric field could drive an insulator to metal transition in a Mott Insulator [3,4,5] appears as particularly appealing. It paves the way for a new type of RRAM based on an Electronic Phase Change, namely a Mott insulator-metal transition with a very fast potential switching time [9].

The $AM_4X_8$ (A = Ga, Ge ; M = V, Nb, Ta, X = S, Se) compounds exhibit a lacunar spinel structure [10] with tetrahedral transition metal clusters $M_4$ (see **Figure S.1** in supporting informations). These compounds are Mott Insulators exhibiting a very small Mott-Hubbard gap (0.2 ± 0.1 eV) due to the presence of $M_4$ clusters [11]. A direct consequence of this low gap value is the high sensitivity to external perturbation such as hydrostatic pressure which can induce an insulator to metal transition in the $AM_4X_8$ compounds [12,13]. In that respect, the $AM_4X_8$ compounds are good candidates to investigate the existence of an electric pulse induce RS in a Mott insulator.

Crystals of $GaV_4S_8$, $GaV_4Se_8$, $GeV_4S_8$, $GaNb_4Se_8$ and $GaTa_4Se_8$ were prepared as explained elsewhere [14]. Two or four gold wires were subsequently connected on these crystals using a carbon paste. All these compounds display an activated behaviour $\rho(T) = \rho_0 e^{E_A/2k_BT}$ at high temperature (T > 200 K), with an activation energy $E_A$ in the 200-330 meV range (see Figure S1). This is consistent with the small Mott-Hubbard gap expected for these five $AM_4X_8$ Mott insulators. We have then applied short electric pulses (10 µs) on these crystals using a simple circuit made of a $AM_4X_8$ crystal in series with a load resistance $R_{load}$ acting as a current limiter after the resistance drop in the $AM_4X_8$ crystal. We found that the non-volatile RS that we had previously observed in $GaTa_4Se_8$ [14,15,16] also occurs in $GaV_4S_8$, $GaV_4Se_8$, $GeV_4S_8$ and $GaNb_4Se_8$. The resistance/resistivity measured for all crystals at low bias level After the Electric Pulse (AEP) reveal a low resistance state with a quasi metallic like behavior at low temperature (see Figure S.1). Below 100 K, the relative ratios $\Delta R/R = (R_{high} - R_{low})/R_{high}$ between the high and low resistance states exceeds 99.9%. At room temperature, $\Delta R/R$ around 5 - 35 % are still observed.

In order to get more insight into the mechanism of the RS, we have monitored the time dependence of the voltage ($V_{sample}$) and current ($I$) across a $GaTa_4Se_8$ crystal during short (10 µs) voltage pulses ($V_{pulse}$) applied to the circuit (see **Figure 1c**). **Figure 1** displays the evolution of the resistance $R(t)$ measured during the pulses and the corresponding $I(V_{sample})$ curve obtained for this crystal. In the pristine high resistance state, a low $V_{pulse}$ of + 25 V keep the resistance unchanged (Fig. 1a(1)); at larger $V_{pulse}$ of + 30 V, a fast (≈ 400 ns) resistance drop occurs at a time $t_{delay} \approx 5$ µs after the beginning of the pulse (Fig. 1a(2)). This transition is purely volatile since the low-bias resistance (indicated by large filled squares in figure 1) measured before and after the voltage pulse are equivalent and remain at a high resistance level. It suggests that the crystal undergoes a fast volatile transition from an insulating to a metallic-like state. A further increase of $V_{pulse}$ up to + 36 V leads to a drop of the resistance in the first 200 ns of the pulse, which turns out to be non-volatile after the end of the pulse (Fig. 1a(3)). Interestingly, the crystal in its low resistance state can switch back to the high resistance state by applying short electric pulses of the same or of reverse polarity. Figure 1d(5) presents the case for which the switch back is obtained with 10 µs pulses of reverse polarity. We note that the pulse inducing the $R_{low} \rightarrow R_{high}$ transition restores a resistance close to the pristine $R_{high}$ value.

Figure 2 displays the resistance $R(t)$ and the voltage $V_{sample}(t)$ measured for successive volatile RS induced to the same $GaTa_4Se_8$ crystal. First we note that the time $t_{delay}$ at which the volatile RS occurs shifts to smaller values when the voltage across the sample increases: $t_{delay} \approx 8.5$ µs for $V_{sample} = 17.3$ V is decreasing to $\approx 5$ µs for 20 V and $\approx 0.2$ µs for 32 V. All these volatile RS are perfectly volatile and are highly reproducible as we checked that $t_{delay}$ remains perfectly constant over several thousand of pulses. As expected, a similar transition is observed for positive and negative electric pulses (*i.e.* this transition is unipolar). Figures 2(b) shows that the Joule energy and the charge released or injected before the transition are not constant for the different pulses. This suggests that the volatile RS does not originate from a trivial Joule or charge injection effect (see figure 2-b). Moreover, the Joule energies are much smaller than 100 nJ which would lead to a maximum temperature increase of 0.1 K in an adiabatic model. On the other hand, figure 2(c) reveals a striking feature : the sample voltage $V_{sample}$ after the RS always fall into the same value $V_{threshold} \approx$ 8-10 V (or $E_{threshold} \approx$ 2-3 kV/cm) regardless the initial voltage experienced by the sample before the RS. Conversely, the current flowing through the sample after the volatile RS varies in a large extent. This corresponds to a highly non-ohmic current-voltage characteristics in the "transited" state, as shown in Fig. 2-d. Interestingly, the electric field in the sample measured just *before* the RS (see inset of Fig. 2-d) extrapolates exactly to the same value $E_{threshold}$ when $t_{delay} \to \infty$. As expected, this $E_{threshold}$ corresponds to the lower voltage inducing a RS in dc measurements. We found that $E_{threshold}$ is of the order of 2-7 kV/cm for all $AM_4X_8$ compounds. Morever, $E_{threshold}$ seems barely dependent on the temperature at which the pulses are applied (77K-150K). Conversely the current density can vary over several order of magnitude in the same temperature range. For example, on $GaNb_4Se_8$ crystals we found $E_{threshold} = 4.3$ kV/cm at 75 K and 4.2 kV/cm at 150 K, while in the mean time the current density was strongly enhanced from 0.5 A/cm$^2$ to 220 A/cm$^2$. These results demonstrate the existence of a threshold electric field $E_{threshold}$ below which no RS can occur in the $AM_4X_8$ compounds. It strongly suggests that the volatile RS is related to an electric-field-driven insulator to "metal" transition.

As already mentioned the RS observed in the $AM_4X_8$ is volatile at low voltage but becomes non-volatile for larger voltage. Analysis of the non volatile transition back to the high resistance state $R_{low} \to R_{high}$ gives some clues about the non-volatile mechanism. The resistance during this pulse (see Figure 1(d-5) for a $GaTa_4Se_8$ crystal) remains in the low resistance state and the resistive switching towards $R_{high}$ occurs systematically *after* the end of the pulse. Moreover, the resistance *during* this pulse (see Figure 1(d-5)) is lower compared to its low bias value measured before the pulse. Similar lowering are always observed for samples in their low resistance state (see *e.g.* figure 2(d-4)). This suggests that the volatile RS mechanism is still operating in the low resistance state and that only a part of the resistance lowering is kept in the non-volatile effect. The non-volatile transition $R_{high} \to R_{low}$ appears then as a two steps mechanism: first an electric-field-driven RS leading to a low resistance state, and secondly a pinning phenomenon which retains a part of the low resistance state. In this scenario, the non-volatile transition $R_{low} \to R_{high}$ would be the consequence of a depinning phenomenon. The microscopic origin of these pinning/depinning phenomena is still not elucidated. However it might be related to the very unusual electrostrictive phenomena (*i.e.* an electric field induced mechanical deformation) observed recently in $GaTa_4Se_8$ [15].

At this step of our analysis, we note that all reported RS mechanisms based on an amorphous-crystalline or a chemical phase change are by nature non-volatile and can not explain the striking volatile RS observed in the $AM_4X_8$ compounds. Also, the volatile and non volatile RS were observed in $GaTa_4Se_8$ and $GeV_4S_8$, which do not share any common chemical element. This clearly invalidates the scenario based on the migration of a specific element. Moreover, the volatile RS is unipolar and can not be explained by a bipolar "electrochemical" mechanism. In the same way, the joule energy released during the pulse is very small, which provides additional evidences against the "thermochemical" and amorphous-crystalline phase-change mechanisms which are based on Joule heating dissipation. Moreover, high resolution analysis performed either by Transmission Electron Microscopy (TEM) or Energy Dispersive X-Rays Diffraction Spectroscopy (EDXS) did not evidenced any local structural or composition change within the sample after a non-volatile transition (see Figure S.2). The RS observed in the $AM_4X_8$ compounds differs therefore from other RS used in PC- or R-RAM and related to an amorphous-crystalline or a chemical phase-change, respectively. Alternatively, all our results demonstrate that the RS observed in the $AM_4X_8$ compounds is related to an electric-field-driven insulator to "metal" transition. It suggests that the electric field can break the fragile Mott insulating state of the $AM_4X_8$ compounds as predicted in recent theoretical works [3,4,5].

The new type of non-volatile RS uncovered here is of great interest as it could be used to build up a new type of RRAM. Figure 3 confirms that the $AM_4X_8$ compounds are able to switch back and forth between the high and low resistance state, at low and at room temperature. This is observed using pulses of the same or of reverse polarity which evidences an unipolar behavior (compare figure 3a and 3b). The relative variation of the resistance *ΔR/R* is very large at low temperature and can reach up to 20 % - 30 % at room temperature. *ΔR/R* values close to 25% as displayed in figure 3d are comparable with those reported on MRAM memories and are therefore compatible with RRAM applications. In $GaTa_4Se_8$ the "transited state" appears quite stable as variations of the resistivity of only a few percent were typically observed after several months at room temperature. This gives hope that the retention time might be good in the $AM_4X_8$ compounds. As shown in figure 3, the RS studied here allows to reach fast writing and erasing times of the order of 50 ns at low temperature and 500 ns at room temperature. This is much better than the writing (1 µs) and erasing times (1 ms) achieved in Flash technology [1]. Another appealing feature of the $AM_4X_8$ lies in the possibility to built up devices operating at voltages lower than 1V as the RS is driven by electric fields of the order of 5kV/cm. This is again much better than the writing/erasing voltages of about 12 V used in Flash memories [1]. All things considered, the $AM_4X_8$ compounds show therefore promising features to build up a new type of non-volatile memory RRAM.

To conclude, we have discovered that the fragile Mott Insulator compounds $AM_4X_8$ exhibit both a volatile and a non-volatile unipolar Resistive Switching (RS). All our results indicate that this RS is related neither to a chemical nor to an amorphous-crystalline phase change as observed in all other systems where a RS occurs. Conversely we have demonstrated that the volatile RS found in the Mott insulators $AM_4X_8$ is related to an electric-field-driven insulator to metal transition. Interestingly, recent theoretical works predict that an electric field can break the Mott insulating state [3,4,5] and induce an insulator to metal transition. The RS observed in the $AM_4X_8$ compounds might be the first example of such a transition. These compounds could therefore offer the possibility to explore a new type of RRAM based mostly on an electronic phase change, that we could call a Mott-RRAM. First results obtained on the $AM_4X_8$ compounds are promising showing fast writing / erasing times (down to 50 ns) and large *ΔR/R* ratio at room temperature (up to 25 %). A long work has now to be engaged to explore further the potential of this new type of RRAM.

*Experimental*

*Transport measurements*: The $AM_4X_8$ crystals used for transport measurements were contacted using 10 µm gold wires and carbon paste (Electrodag PR-406), and then annealed in vacuum at 150°C during 30 minutes. The low-bias resistance of the $AM_4X_8$ was measured using a Keithley 6430 source-measure unit by a standard two- or four-probe technique. We checked that the contact resistances were much smaller than the sample resistance. Voltage pulses were applied using an Agilent 8114A. During the pulse, the voltage and current across the sample were measured with a Tektronix DPO3034 oscilloscope associated with a IeS-ISSD210 differential probe, using the simple circuit described in Fig 1-c.

*Acknowledgements*

The authors thank J. Martial for sample preparation, and E. Souchier, M.-P. Besland, P. Moreau and V. Fernandez for their commitment in the project. This work was supported by the French Agence Nationale de la Recherche through the funding of the "NV-CER" (ANR-05-JCJC-0123-01) and "NanoMott" (NT09_513924) projects.

Received: ((will be filled in by the editorial staff))
Revised: ((will be filled in by the editorial staff))
Published online: ((will be filled in by the editorial staff))

**Figure 1**: time dependence of the electrical resistance of a $GaTa_4Se_8$ single crystal measured between (large filled squares) and during (a-1 to a-3 and d-4 to d-6) the application of electric pulses at 77 K. The voltage of each pulse $V_{pulse}$ applied on a the crystal in series with a load resistance of 20 kΩ is indicated above or below each $R(t)$ graph. The small curvature of $R(t)$ observed in the beginning of the pulses are related to extrinsic capacitive effects. The current *vs.* voltage (electric field) curves of this sample (b) are measured during the pulse by monitoring the voltages $V_{load}$ and $V_{sample}$ as shown in (c). Open circles in R(t) graphs (a) and (d) indicate the sample current and voltage data reported in (b)

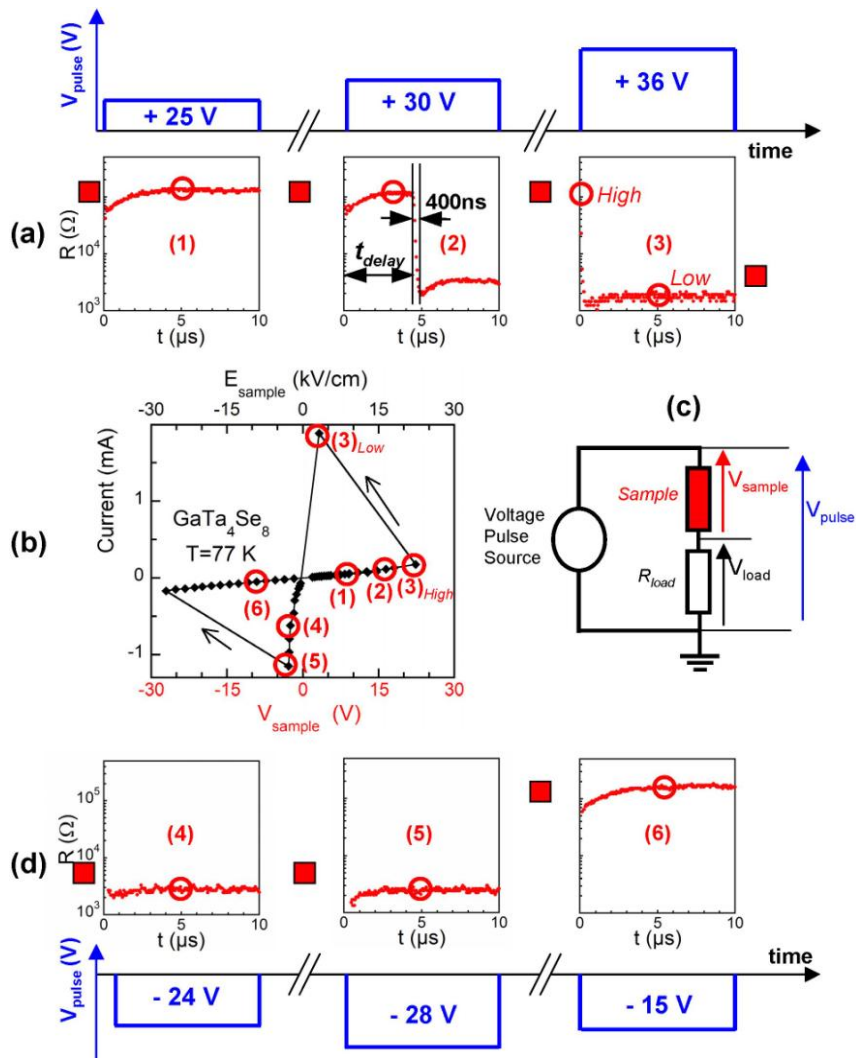

**Figure 2**: (a) time dependence of the resistance *R(t)* of a GaTa$_4$Se$_8$ single crystal during the 10 µs pulse for several voltages applied at 77 K. (b) Joule energy $E_J$ and injected charge Q calculated as $E_J = \int_0^{t_{delay}} P(t)dt = \int_0^{t_{delay}} V_{sample} V_{load}/R_{load} dt$ and $Q = \int_0^{t_{delay}} I dt = \int_0^{t_{delay}} V_{load}/R_{load} dt$ released in the crystal between the beginning of the pulse and the time $t_{delay}$ at which the volatile transition occurs. (c) voltage $V_{sample}(t)$ across the sample, same data and color code as in (a). (d) current-voltage characteristics measured *during* the pulses, before (black circles) and after (filled squares) the volatile transition. Inset : sample voltage before and after RS *vs* $(1/t_{delay})^{1/3}$. The exponent α = 1/3 is arbitrarily chosen to give a linear dependence of $V_{sample}$ *vs* $(1/t_{delay})^\alpha$.

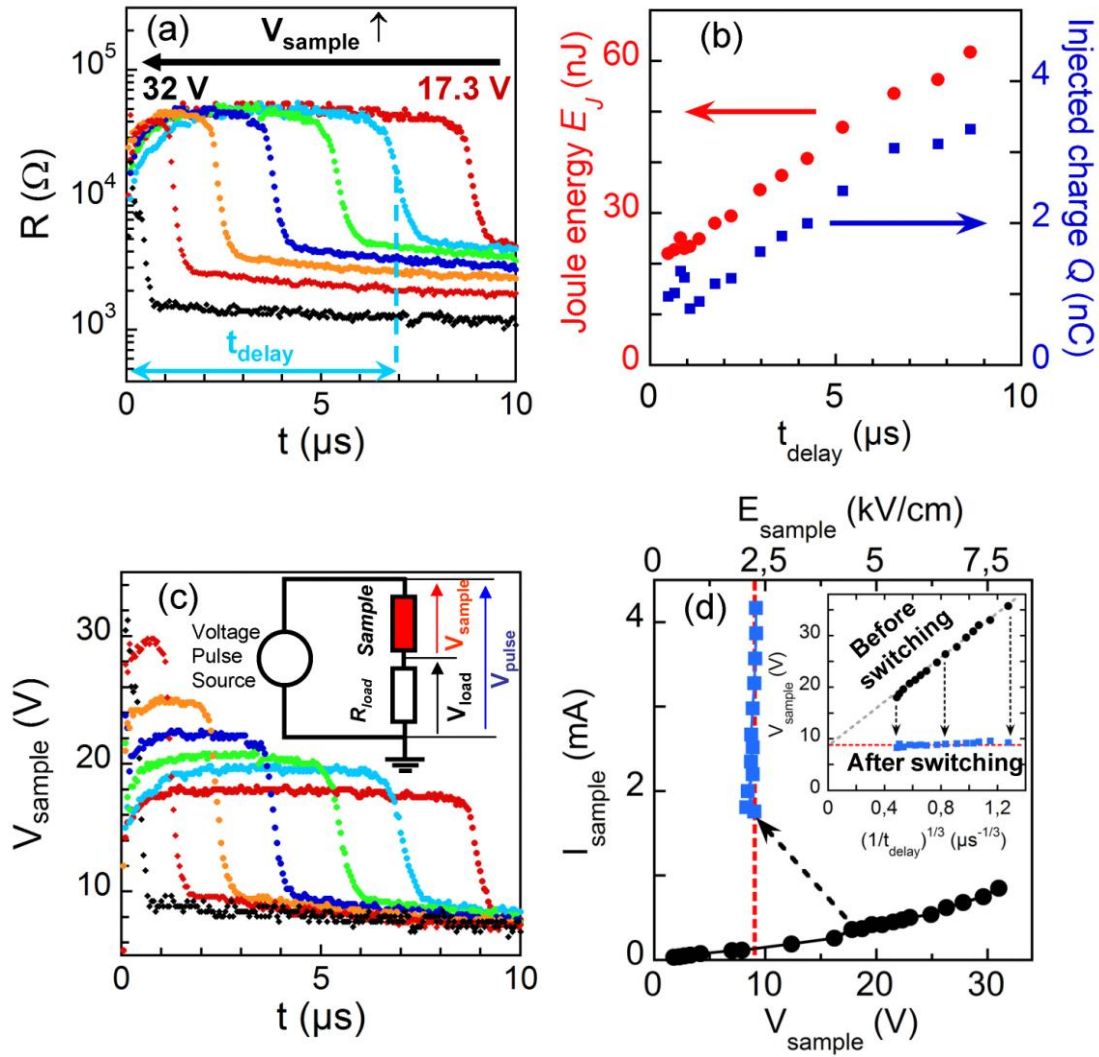

**Figure 3**: switching performances of a GaV$_4$S$_8$ single crystal obtained at 90 K with series of 200 ns pulses of the same polarity (a), at 90 K with 50 ns pulses of alternating polarity (b), at room temperature with 500 ns (d) and 10 μs (e) pulses of the same polarity. The crystal was connected in the "capacitive" geometry, with an inter-electrode distance of about 40 μm.

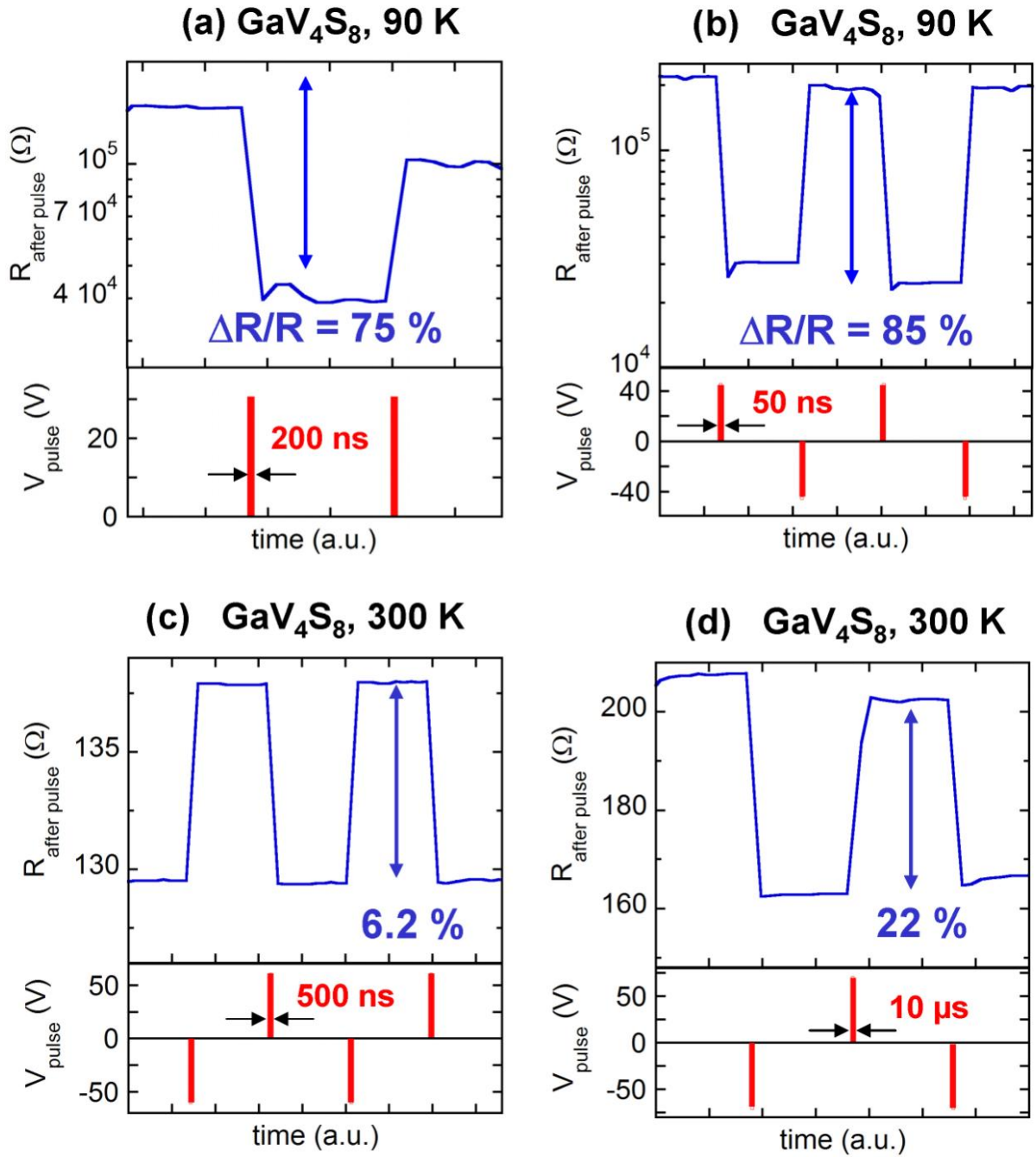

**Table of contents**

**Electric-Field-Induced Resistive Switching in a Family of Mott Insulators : towards Non-Volatile Mott-RRAM Memories**

Laurent Cario, Cristian Vaju, Benoit Corraze, Vincent Guiot, Etienne Janod*

Flash memories (USB keys) are close to their miniaturization limits. The ultimate evolution of such devices is believed to exploit different concepts such as electronic phase transitions. Here we show that electric field can trigger fast resistive switching in the fragile Mott insulators $AM_4X_8$. This could be the first example of such new non-volatile memories : the Mott-RRAM.

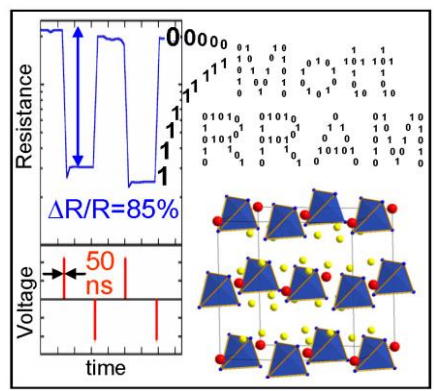

# Supporting Informations

**Figure S1.** (a) crystallographic structure of the $AM_4X_8$ compounds highlighting the $M_4$ tetrahedra, the $AX_4$ tetrahedra and the $MX_6$ octahedra. (b-f) temperature dependence of the resistance or resistivity of the different $AM_4X_8$ compounds in their pristine high resistance state and in their electric-pulse-induced low resistance state. For all samples electric pulses were applied around 100K. Insets indicate the structure of the molecular orbitals associated with the $M_4$ clusters.

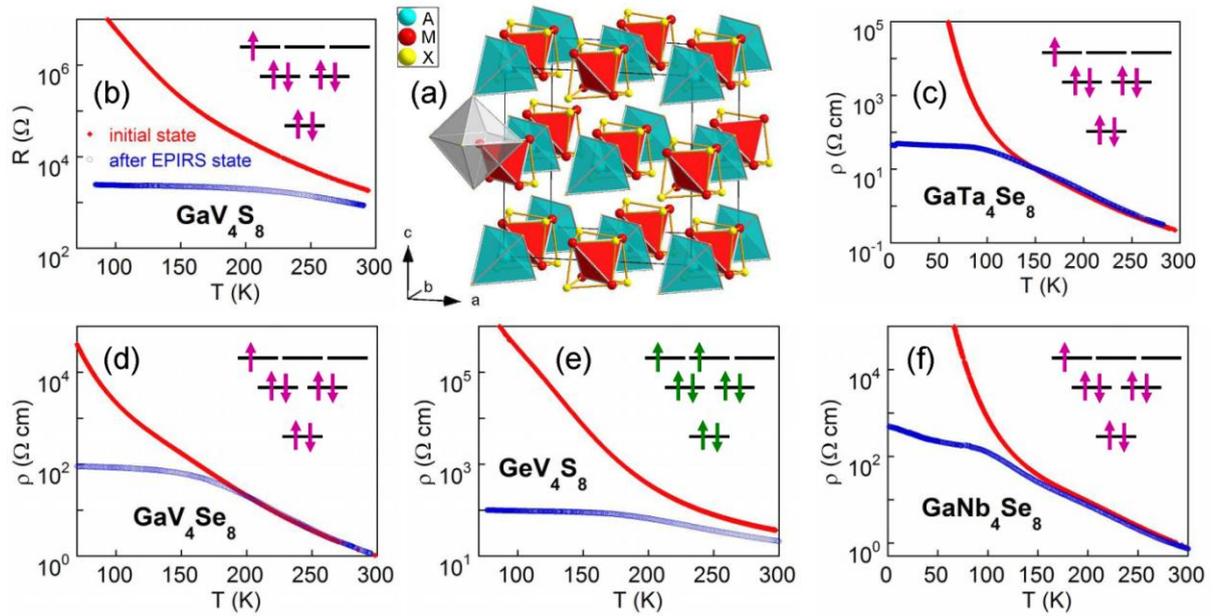

**Figure S2 :** typical high resolution TEM image of GaV$_4$S$_8$ in its pristine state. Calculated images performed with the JEMS program and direct representations of its crystallographic structure (space group $F\bar{4}3m$) appear as insets. Our systematic TEM analysis of AM$_4$X$_8$ single crystals in their pristine state and after an electric-pulse-induced resistive switching did not allow detecting any obvious modifications such as local amorphisation, distorsion or phase transition.

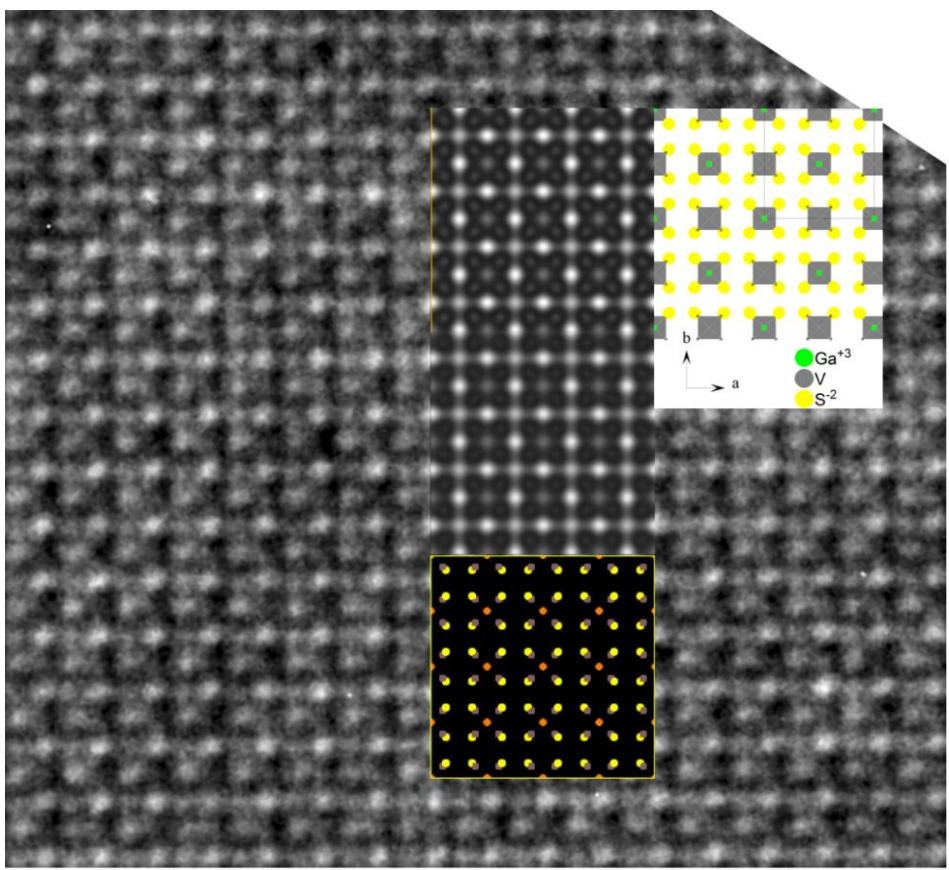